\documentclass{pnastwo}

\usepackage{graphicx}
\usepackage{amssymb,amsfonts,amsmath}
\usepackage[normalem]{ulem}

\makeatletter
\long\def\@makecaption#1#2{%
\ifx\@captype\table
\let\currtabcaption\relax
\gdef\currtabcaption{
\tabnumfont\relax #1. \tabtextfont\relax#2\par
\vskip\belowcaptionskip 
}
\else
 \vskip\abovecaptionskip
  \sbox\@tempboxa{\fignumfont#1.\figtextfont\hskip.5em\relax #2}%
  \ifdim \wd\@tempboxa >\hsize
\fignumfont\relax #1.\figtextfont\hskip.5em\relax#2\par
  \else
    \global \@minipagefalse
    \hb@xt@\hsize{\hfil\box\@tempboxa\hfil}%
  \fi
\fi
}
\makeatother

\makeatletter                                  
\def\DonormalEndcol{%
\ifx\toporbotfloat\xtopfloat%
  \ifcaptypefig%
  \expandafter\gdef\csname topfloat\the\figandtabnumber\endcsname{%
  \vbox{\vskip\PushOneColTopFig%
  \unvbox\csname figandtabbox\the\loopnum\endcsname%
  \vskip\abovefigcaptionskip%
  \csname caption\the\loopnum\endcsname%
  \csname letteredcaption\the\loopnum\endcsname%
  \csname continuedcaption\the\loopnum\endcsname%
  \csname letteredcontcaption\the\loopnum\endcsname            
  \ifredefining%
  \csname label\the\loopnum\endcsname%
  \expandafter\gdef\csname topfloat\the\loopnum\endcsname{}\fi}%
  \vskip\intextfloatskip
  \vskip-4pt 
}%
\else%
  \ifcaptypeplate%
  \expandafter\gdef\csname topfloat\the\figandtabnumber\endcsname{%
  \vbox{\vskip\PushOneColTopFig%
  \unvbox\csname figandtabbox\the\loopnum\endcsname            
  \vskip\abovefigcaptionskip                           
  \csname caption\the\loopnum\endcsname                    
  \csname letteredcaption\the\loopnum\endcsname                
  \csname continuedcaption\the\loopnum\endcsname               
  \csname letteredcontcaption\the\loopnum\endcsname            
  \ifredefining                                
  \csname label\the\loopnum\endcsname                      
  \expandafter\gdef\csname topfloat\the\loopnum\endcsname{}\fi}        
  \vskip\intextfloatskip 
  \vskip-4pt 
}%
\else
 \expandafter\gdef\csname topfloat\the\figandtabnumber\endcsname{%
 \vbox{\vskip\PushOneColTopTab 
 \csname caption\the\loopnum\endcsname                     
  \csname letteredcaption\the\loopnum\endcsname                
  \csname continuedcaption\the\loopnum\endcsname               
  \csname letteredcontcaption\the\loopnum\endcsname            
  \vskip\captionskip                               
  \unvbox\csname figandtabbox\the\loopnum\endcsname            
\ifredefining                                  
\csname label\the\loopnum\endcsname                    
\expandafter\gdef\csname topfloat\the\loopnum\endcsname{}\fi           
}\vskip\intextfloatskip 
\vskip-10pt}                                   
\fi\fi%
\else
\ifcaptypefig                                  
\expandafter\gdef\csname botfloat\the\figandtabnumber\endcsname{%
\vskip\intextfloatskip                             
\vbox{\unvbox\csname figandtabbox\the\loopnum\endcsname            
\vskip\abovefigcaptionskip                         
  \csname caption\the\loopnum\endcsname                    
  \csname letteredcaption\the\loopnum\endcsname%
  \csname continuedcaption\the\loopnum\endcsname%
  \csname letteredcontcaption\the\loopnum\endcsname%
\vskip\PushOneColBotFig
\ifredefining%
\csname label\the\loopnum\endcsname                    
\expandafter\gdef\csname botfloat\the\loopnum\endcsname{}\fi}}%
\else                                      
\ifcaptypeplate                                
\expandafter\gdef\csname botfloat\the\figandtabnumber\endcsname{%
\vskip\intextfloatskip                             
\vbox{\unvbox\csname figandtabbox\the\loopnum\endcsname            
\vskip\abovefigcaptionskip                         
  \csname caption\the\loopnum\endcsname                    
  \csname letteredcaption\the\loopnum\endcsname%
  \csname continuedcaption\the\loopnum\endcsname%
  \csname letteredcontcaption\the\loopnum\endcsname%
\vskip\PushOneColBotFig
\ifredefining%
\csname label\the\loopnum\endcsname                    
\expandafter\gdef\csname botfloat\the\loopnum\endcsname{}\fi}}%
  \else
\expandafter\gdef\csname botfloat\the\figandtabnumber\endcsname{%
  \vskip\intextfloatskip                           
\vbox{\csname caption\the\loopnum\endcsname                
  \csname letteredcaption\the\loopnum\endcsname                
  \csname continuedcaption\the\loopnum\endcsname               
  \csname letteredcontcaption\the\loopnum\endcsname%
  \vskip.5\intextfloatskip                         
  \unvbox\csname figandtabbox\the\loopnum\endcsname%
\vskip\PushOneColBotTab                            
\ifredefining%
\csname label\the\loopnum\endcsname                    
\expandafter\gdef\csname botfloat\the\loopnum\endcsname{}\fi}}%
\fi\fi\fi}                                 
\makeatother

\url{www.pnas.org/cgi/doi/10.1073/pnas.0709640104}
\copyrightyear{2008}
\issuedate{Issue Date}
\volume{Volume}
\issuenumber{Issue Number}
\begin{document}


\title{Real Space Renormalization Group Theory of Disordered Models of Glasses}

\author{
Maria Chiara Angelini\affil{1}{Dipartimento di Fisica, Ed. Marconi, "Sapienza" Universit\`a di Roma, P.le A. Moro 5, 00185 Roma Italy}
Giulio Biroli\affil{2}{Institut Physique Th\'eorique (IPhT) CEA Saclay, and CNRS URA 2306, 91191 Gif Sur Yvette, }
\affil{3}{Laboratoire de Physique Statistique, Ecole Normale Sup\'erieure, PSL Research University; Universit\'e Paris Diderot Sorbonne Paris-Cit\'e; Sorbonne Universit\'es UPMC Univ. Paris 06; CNRS; 24 rue Lhomond, 75005 Paris, France.}}

\contributor{Submitted to Proceedings of the National Academy of Sciences
of the United States of America}

\significancetext{The hallmark of the glass transition is the very fast increase of the relaxation time $\tau_\alpha$. Experimental results 
have rationalised this fact in terms of a growing effective activation energy scale, $\Delta=T\log \tau_\alpha$, characterising the height of the barriers in the rough energy landscape of glass-forming liquids. By applying the renormalisation group method to
disordered models of glasses we show that barriers related to degrees of freedom correlated on the scale $\ell$ 
scale as  $\ell^\theta$ where $\theta$ is a critical exponent. The growth of the static length-scale $\xi$,  
which is the typical size of the statically correlated regions, thus leads to  $\Delta \sim \xi^\theta$ and 
provides a general explanation for glassy behaviour.     
}
\maketitle

\begin{article}
\begin{abstract}
We develop a real space renormalisation group analysis of disordered models of glasses, in particular of the spin models at the origin of the Random First Order Transition theory. We find three fixed points respectively associated to the liquid state, to the critical behavior and to the glass state. The latter two are zero-temperature ones; this provides a natural explanation of the growth of effective activation energy scale and the concomitant huge increase of relaxation time
approaching the glass transition. The lower critical dimension depends on the nature of the interacting degrees of freedom and is higher than three for all models. This does not prevent three dimensional systems from being glassy. Indeed, we find that 
their renormalisation group flow is affected by the fixed points existing in higher dimension and in consequence is non-trivial. Within our theoretical framework the glass transition results to be an avoided phase transition. 
\end{abstract}
\vspace{-.5cm}
\section{Introduction}
Understanding the glass transition and the properties of glasses is still an open problem despite many years of research and a lot of recent progress \cite{rmp,physicstoday}. 
The glass state poses to statistical physicists a conundrum and a challenge. It is a state of matter that is rigid like a crystal but 
disordered like a liquid. A glass is a solid that forms progressively from a super-cooled liquid in a way 
strongly suggesting the existence of an underlying critical phenomenon. In fact, the relaxation time-scale and the viscosity of super-cooled liquid increase very fast when lowering the temperature, indicating a growing effective energy scale as it would be expected at a phase transition. If such a phase transition really exists (or at least is weakly avoided), then it certainly has to be of a new kind since the  corresponding phenomenology does not fit in any standard paradigm. The challenge in order to understand glasses and the glass transition is that one has to deal with a system of many interacting degrees of freedom characterised by a very rugged energy landscape. This is a fascinating problem that has ramifications in contexts as different as computer science and biology as well as in other branches of physics.  Standard tools of statistical mechanics are not well adapted to tackle it. In fact, part of the current research is devoted to develop new theoretical methods and concepts able to deal with it \cite{perspective}.\\
Among the several different theoretical approaches aimed at explaining the glass transition, the Random First Order Transition theory 
originally introduced by Kirkpatrick, Thirumalai and Wolynes (KTW) stands as one of the most prominent candidate \cite{RFOT1,RFOT2}. In a pioneering 
work, KTW realised that the physics of super-cooled liquids approaching the glass transition is remarkably similar to the one 
of generalised mean-field spin-glass models and used this as a starting point to develop what is now called RFOT theory. 
What at that time was considered by some a great insight (but also an absurd assumption by others) proved to be right in recent years. 
It is now clear that the mean-field theory of glasses, defined in terms of the solution in the limit of large spatial dimension $d$, is 
the same of the generalised spin-glass models initially considered by KTW. This result comes from a long series of works that culminated in the exact solution of hard spheres in the infinite dimensional limits \cite{HSHighd}. The solution of models with smooth potentials at finite temperature 
was also completed recently \cite{Pierf} (see also \cite{jorgeFZ}). Going beyond $d=\infty$ results is a quite challenging task and it is at the core of current research: 
A phenomenological scaling theory for finite dimensional systems was originally developed by KTW \cite{RFOT1} and in recent years theoretical and numerical results have developed further 
these ideas \cite{BB,Dzero,Franz,BC}.  What makes going beyond mean-field theory hard is that the growth of time and length-scales approaching the glass transition is related to non-perturbative effects in $1/d$. Unlike in standard critical phenomena, where mean-field theory is quantitatively correct above the upper critical dimension and qualitatively correct below, for glasses mean-field theory has many caveats even in large dimensions, the most problematic being the infinite life-time of metastable amorphous states (barriers diverge exponentially in $d$). The very process 
characterising slow and glassy dynamics--the evolution from one metastable state to another--is absent in mean-field theory, which therefore cannot explain by itself the growth of time and length-scale approaching the glass transition. \\
In standard critical phenomena the Renormalisation Group (RG) proved to be the theoretical framework that allows one to go beyond the mean-field approximation and obtain a complete theory. In the case of systems characterised by a rugged energy landscape, 
such a method is still far from being applicable in general. There are both technical and conceptual difficulties. On the one hand, in this case it is necessary to develop an RG procedure able to take into account the complexity of the rugged landscape: one cannot follow the flow of a few coupling constants but instead the RG has to be functional and in many cases non-perturbative. On the other hand, fixed points describing the critical behaviour and the low temperature phase are possibly of new kinds and require a conceptual leap forward.  
In some cases, such as Random Manifolds \cite{Fisher,Wiese} and the Random Field Ising Model  \cite{RFIM_NPRG} (RFIM), these difficulties were overcome and a complete theory was worked out; in others, such as glasses, the state of the art is less advanced despite some recent progress \cite{moore,castellana,CBTT}. \\
With the aim of developing an RG analysis of the glass transition, in this work we will present a complete real space RG analysis of the class of disordered models from which RFOT originated. 
Since at least within mean-field theory spins and particles models are in the same universality class, it is conceivable that our findings could hold for particle systems too.   
Before summarising our main results, it is useful to discuss in general the kind of RG flow one expects for a system undergoing a glass transition. Several ideas were already put forward in the literature \cite{RFOT1,CBTT,sethna,mooredrossel,KCMRG}. In the following we will take the point of the view which advocates  that the origin of the dynamical slowing down lies in the thermodynamics and that in consequence understanding the properties of the landscape and static correlations is the key ingredient to develop a theory of the glass transition. We will therefore focus on the thermodynamics and infer how this influences dynamics. This assumption is rooted in the RFOT theory and also in other approaches \cite{RFOT2,FLD}; it received positive but not conclusive confirmations recently (see \cite{yaida1,yaida2} and refs therein).\\ The very basic experimental fact about systems approaching the glass transition is the dramatic increase of the relaxation time $\tau_\alpha$: 
the effective energy scale (henceforth denoted $\Delta$) increases, possibly as a diverging power law in the distance from the transition, and leads to a growth of $\tau_\alpha$ which is much faster than the usual power law of standard critical phenomena. This strongly suggests that the critical point associated to the glass transition (if it exists) should be a zero-temperature one \cite{fisherRFIM}, meaning that $\Delta(\ell)$ should scale as $\Delta(\ell)\simeq \Delta_0 \ell ^\theta$ where $\ell$ is the 
scale over which fluctuations have been integrated out by RG. Close to the critical point, under renormalisation, the system is 
characterised by a larger and larger $\Delta$ until the static correlation length $\xi$ is reached. On scales larger than $\xi$ the system can be considered as an ensemble of weakly interacting sub-parts, each one characterised by the energy scale $
\Delta (\xi)$. The time-scale for relaxation is therefore given by the Arrhenius law applied to each sub-part. This leads to $\tau\simeq \tau_0e^{\Delta(\xi)/T}=\tau_0e^{\Delta_0 \xi^\theta/T}$, which is the law conjectured to hold (for different reasons) in the different thermodynamic theories
of the glass transition \cite{RFOT2,FLD}. This gedanken description of the RG flow for glasses is simplified in many ways. Among the several possible additional phenomena that could have been taken into account, three are particularly important. First, as discussed in the introduction, rugged landscapes are 
complex object that cannot be simply described by just one coupling constant; one has at least to consider that $\Delta$ fluctuates in space
so what really flows is the distribution $P(\Delta)$. In consequence, in the discussion above, $\Delta$ should be considered the typical value of the energy scale extracted from  $P(\Delta)$.
This difficulty is related to an important physical ingredient: glassy systems are disordered because they contain either quenched or self-induced disorder. 
Disorder leads naturally to fluctuations of the energy scales and also to a competition between the energy scale 
characterising the interaction between the degrees of freedom, $\Delta$, and an idiosyncratic energy scale, that we shall denote $h$, favouring specific local configurations (In the case of the RFIM the former is the ferromagnetic coupling and the latter is the random field).
Both the distribution of $\Delta$ and $h$ have to be taken into account, since they clearly compete one with the other. The idiosyncratic disorder $h$ could even preclude the growth of $\Delta$ and, hence, wipe out the transition. 
Our final caution is that the critical point discussed above might be avoided but still be meaningful. In this case, when lowering the temperature the RG trajectory could approach the critical fixed point and, at lower temperature, the fixed point describing the solid phase but without never flowing onto it. This is the situation advocated for the frustration limited domain theory \cite{FLD}, that could also happen for RFOT \cite{RFOT2}: The ideal glass transition does not really takes place for physical systems but it conceivably does for a modified version of the system (e.g. in curved space or higher dimension); nevertheless, if the regime where the critical point is avoided is experimentally inaccessible, for all practical purposes the fixed points identified by the RG flow (in a modified version of the system) govern glass transition physics.   \\
\vspace{-1.3cm}
\begin{figure}[h]
\begin{center}
 \includegraphics[width=0.28\textwidth]{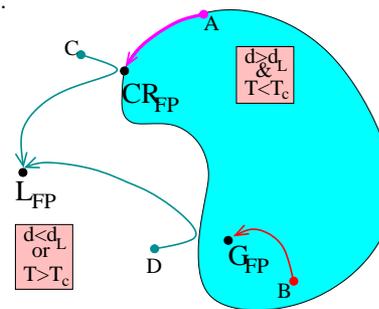}
\end{center}
\caption{The infinite dimensional space of coupling constants is 
represented in two dimensions by a cartoon.  The two stable fixed points,  $\text{G}_{\text{FP}}$ and $\text{L}_{\text{FP}}$, 
correspond respectively to the glass and liquid phase. The light blue and white regions correspond to their basins of attraction.
On the boundary between them lies the critical fixed point $CR_{FP}$. 
The initial condition for the flow depends on the temperature, the model and the spatial dimension: for $d\ge d_L$ at $T=T_c $ 
the system lie on the critical manifold and is attracted by $CR_{FP}$ (point A), whereas for $T<T_c$ falls onto $G_{FP}$ (point B); a model close to criticality, 
is initially attracted by $CR_{FP}$ but it eventually bifurcates toward $L_{FP}$, this can happen either below or above $d_L$ (point C); 
for $d<d_L$ and $T<<T_c^{d_L}$ the flow is attracted by $G_{FP}$ but it eventually bifurcates toward $L_{FP}$, this leads to 
avoided criticality (point D). 
}
\label{Fig:Cartoon}
\end{figure}
\vspace{-.3cm}
What we find in our analysis contains all the ingredients discussed above. At high enough dimension, higher than three, disordered models of glasses are characterised by a zero temperature critical fixed point, $CR_{FP}$, that 
emerges from the competition between $\Delta$ and $h$ and governs the critical behaviour associated to the glass transition. We also find 
a non-trivial fixed point, $G_{FP}$, describing the low temperature phase, i.e. the amorphous solid state, to which the system flows on large length-scales for $T<T_c$. Some models, characterised by a too strong idiosyncratic disorder, lie in RG sense far from the basin of attraction of  $G_{FP}$ and, hence, they do not display glass behaviour in finite dimension. Others are instead characterised by a bona-fide glass transition. By studying the dependence of  $CR_{FP}$ and $G_{FP}$ on dimensionality, our results indicate that both fixed points simultaneously disappear without coalescing for a dimension $d_L$ higher than three. All that is summarised pictorially in Fig.1.
Note that the larger is the number of degrees of freedom, $2^M$, strongly interacting on small length-scale, the lower is $d_L$ and the closer approaches three but never reaches it. By considering also non-integer dimensions, as usually done in RG study, we find that $d_L(M)\rightarrow 4.18$ for $M\rightarrow \infty$. 
Due to the regularity of the RG flow, the behavior below $d_L$ is very similar to the one at $d_L$; it departs from it only when 
approaching the vestige of   $CR_{FP}$ and $G_{FP}$ (the region in couplings space where $CR_{FP}$ and $G_{FP}$ lie for $d=d_L$). 
This happens on larger and larger length-scale the more $d$ is close to $d_L$. In fact, by analysing the RG flow in three dimension, we find that although the glass transition is avoided, the existence of fixed points at $d_L$ affects the flow for $d=3$ and leads to glassy behaviour and avoided criticality. 
 \section{Analyzed models}
Our analysis applies to generic spin or Potts disordered finite-range models defined on a cubic (or hyper-cubic) lattices. Since the interaction 
range is finite, i.e. degrees of freedom distant more than $\ell_{I}$ do not interact, it is useful to divide the lattice in a sub-lattice of blocks of size $\ell_{I}^d$. The degrees of freedom inside one block can be "packed" into one single degree of freedom $\sigma_i$ that take $q=2^M$ values, where e.g. $M={\ell_{I}^d}$  for spin  variables. In terms of these new variables 
$\sigma_i$ we have now a model with interaction between closest neighbours only. Without loss of generality, we consider only pairwise 
interactions between the $\sigma_i$ of the closest blocks \footnote{There could be also multi-$\sigma_i$ interactions between closest blocks. We neglect them for simplicity. In any case, they become negligible with respect to the pairwise interaction between nearest neighbour blocks 
if the size of the blocks is taken much larger than $\ell_{I}$.}. We therefore end up with a very generic set of models, that we will call the $M-$value models,  which 
are nearest neighbor spin disordered models for which "spins" can take $q=2^M$ values and with a Hamiltonian of the form:
\begin{equation}
H(\{\sigma\})=\sum_{(i,j)}E_{i,j}(\sigma_i, \sigma_j).
\label{eq:H}
\end{equation}
where $E_{i,j}(\sigma_i, \sigma_j)$ are a set of random link-energies. Depending on their choice one can realise a large variety of models.  
In particular all disordered spin and Potts systems known to have a RFOT, and thus glassy features, within the mean-field approximation belong to this general class. The RG analysis of the M-value models thus allows us to draw very general conclusions. Different models characterised by the same value of $M$
only correspond to different initial condition for the RG flow, but the fixed points  that they can possibly reach are the same; they only depend on the value of $M$.
 \\
The models we have focused on in this work are: 
\begin{itemize}
\item {\it Finite Dimensional Random Energy Models (REM)}. 
The REM is the simplest mean-field model of glasses \cite{derrida,grossmezard}. Its finite dimensional version 
is defined by  taking all $E_{i,j}(\sigma_i, \sigma_j)$ in eq. (\ref{eq:H}) as independent identically distributed (i.i.d.) Gaussian random variables extracted with zero mean and variance $M/2$. 
This system, defined by eq. (\ref{eq:H}), is then a collection of coupled finite size REMs and corresponds to 
a Kac version of the REM \cite{REM1D}. It has been studied  as a function of $M$ in one dimension \cite{REM1D}:
of course no phase transition is present in this case, but the behaviour on finite length-scales approaches (very) slowly the mean-field one in the 
limit $M\rightarrow \infty$.   
\item {\it $M-p$ spins disordered models}. 
These are finite-dimensional versions of the p-spin disordered models that have been extensively studied as mean-field 
models of glasses: each site hosts M Ising spins, hence $\sigma$ has $2^M$ different values. 
The interaction between the spins is p-body, random and involves the subspins belonging to two sites which are neighbours on the cubic (or hyper-cubic) lattice, 
see \cite{Mp}. 
This model has an RFOT transition in its mean-field version for large enough values of $M>M_c(p)$ \cite{Mp,CBTT2}. 
\item {\it Super-Potts Models}. 
In this case on each site there is a Potts variable characterised by $q=2^M$ "colors" and 
 $ E_{i,j}(\sigma_i, \sigma_j)=E_0$ for $(\sigma_i, \sigma_j)=(\sigma_i^*, \sigma_j^*)$ and 
$ E_{i,j}(\sigma_i, \sigma_j)=E_1$ otherwise; $(\sigma_i^*, \sigma_j^*)$ is randomly drawn among the 
$2^M\times 2^M$ possible couples $(\sigma_i,\sigma_j)$ independently for any couple of neighbors $(i,j)$ \cite{SP}. These systems have been shown analytically 
to display an RFOT transition within mean-field theory for $2^M$ larger than $20$ and numerically to display evidences of a glass transition in three dimensions for $2^M$ larger than $30$ \cite{SP}. 
Super-Potts models are generalisations of the disordered Potts glasses \cite{sompo} originally considered by KTW. 
We introduced them in order to bypass the problems (weak frustration and lack of glassy behaviour) found for disordered Potts model in three dimension.
\end{itemize}
Several other relevant models belong to the $M$-class, in particular the random permutation Potts glasses \cite{PermutationMarinari1,PermutationMarinari2} and the third nearest neighbours 
disordered Potts model recently introduced \cite{giap}.
Our analysis can be easily extended to those. 
\section{Nature of the couplings and RG procedure}
In order to gain some general intuition about the model defined in eq. (1) let us decompose the couplings in the following way: 
\begin{equation}
E(\sigma_1,\sigma_2)=-J(\sigma_1,\sigma_2)-V_L(\sigma_1)-V_R(\sigma_2)+C.
\label{Eq:Er}
\end{equation}
where $J$ and $V$ and $C$ can be interpreted respectively as the true interactions between degrees of freedom, the external fields and a constant. They can be fixed univocally from $E(\sigma_1,\sigma_2)$ using the following set of relations. $C$ is naturally defined via the equation $$C=\frac{1}{2^{2M}}\sum_{\sigma_1,\sigma_2}E(\sigma_1,\sigma_2)$$ 
The relation verified by the external field contribution on $\sigma_1$ ($\sigma_2$) is $\sum_{\sigma_1}V_L(\sigma_1)=0$ ($\sum_{\sigma_2}V_R(\sigma_2)=0$). It encodes the fact that the field should favour configurations of $\sigma_1$ independently of the value of $\sigma_2$, hence by summing over $\sigma_1$ one must get a constant, which has to be taken equal to zero to avoid double-counting with $C$. 
Finally, the interaction is what remains of $E(\sigma_1,\sigma_2)$ when the field and the constant are subtracted off and is defined via the relation
 $\sum_{\sigma_1}J(\sigma_1,\sigma_2)=\sum_{\sigma_2}J(\sigma_1,\sigma_2)=0$. 
 By inverting these equations one finds:
 \[
 V_R(\sigma_2)=-\frac{1}{2^M}\sum_\tau E(\tau,\sigma_2)+C \,;\, V_L(\sigma_1)=-\frac{1}{2^M}\sum_\tau E(\sigma_1,\tau)+C
 \]
  \[
 J(\sigma_1,\sigma_2)=-E(\sigma_1,\sigma_2)+\frac{1}{2^M}\sum_\tau E(\tau,\sigma_2)+\frac{1}{2^M}\sum_\tau E(\sigma_1,\tau)-C
 \]
 It is easy to verify that for standard spin models these equations lead to the usual definition of magnetic interaction and magnetic field. Note that even if some of those terms are not present in the original model they are generated by the 
 RG flow. \\
 The decomposition performed in (2) makes clear that the models we are considering can be thought as a generalised version of 
 random magnets with pairwise interactions. A natural method to perform a real-space RG procedure is therefore using the Migdal-Kadanoff (MK) approximation \cite{MK}. 
 This was proven to be able to capture non-perturbative zero temperature fixed points, such as the ones conjectured to play a role for the glass transition, and therefore 
 it is a very good candidate to perform an RG analysis. In the case of 
 spin-glasses MK-RG provides a basis for the droplet model and correctly finds that the critical point is standard whereas 
 the low temperature fixed point is non-trivial \cite{SG_MK}. In the case of the random field Ising model it correctly finds that the critical behaviour 
 is governed by a zero temperature fixed point whereas the low temperature phase is trivially ferromagnetic \cite{RFIM_MK}. MK-RG has other advantages,
 in particular it is exact in one dimension and often provides a good quantitative approximation for the values of the critical exponents in not too high dimensions. Needless to say, it also suffers of some important drawbacks, in particular it does not reveal the existence of an upper critical dimension (when this exists) and does not allow to make a direct connection with mean-field theory. All in all, however, it has proven 
 to provide valuable guidelines for the behaviour of finite dimensional systems. It has a predictive power similar to the one of mean-field theory for phase diagrams: it provides a qualitative correct description of RG flows, very often describing correctly the nature of the fixed points but being unable to provide accurate results except in low dimensions. In the following we present the analysis of the RG flow
of the models (1) by MK-RG.\\
There are several variants of the MK-RG. We use the one that turns out to be exact on hierarchical lattices (HL).  
A HL is generated iteratively, see the inset of Fig. 2: at the step $G=0$ two spins are connected by a single link. 
At each step $G$, each link is replaced by $b$ parallel branches, made of $2$ bonds. 
The effective dimension of this model is $d=1+\ln(b)/\ln(2)$ \cite{MK,SG_MK,RFIM_MK}.
The MK-RG, which is exact on HL and only approximate for hypercubic lattices in finite dimensions,
consists in integrating out progressively degrees of freedom on larger and larger scales. 
It proceeds by retracing back the construction of the HL and iteratively summing over the $\sigma$ on each branch.  
If we call $\sigma_3^i$, the spin traced out in the branch $i$, 
the new renormalized energy $E^R_{1,2}$ between the two external spins $\sigma_1$ and $\sigma_2$ reads:
\begin{equation}
e^{-\beta E^R_{1,2}(\sigma_1,\sigma_2)}\equiv \prod_{i=1}^b\left(\sum_{\sigma_3^i=1}^{2^M}e^{-\beta \left((E_{1,3}^i(\sigma_1,\sigma_3^i)+E_{3,2}^i(\sigma_3^i,\sigma_2)\right))}\right).
\nonumber
\end{equation}
In this way after one RG-step we get a renormalised M-value model whose unit of length is doubled.    
The previous equation defines the RG procedure for the probability distribution of $E^R_{1,2}(\sigma_1,\sigma_2)$ 
and allow us to obtain its RG flow over larger and larger length-scales. 
There is an additional rule which is needed in cases, such as the one we are considering, 
in which external fields are present or generated: before summing over the internal spins, the fields contribution to the couplings ${E}_{1,3}^i(\sigma_1,\sigma_3^i)$ and $E_{3,2}^i(\sigma_3^i,\sigma_2)$ are moved 
from the external spins to the internal ones for all but one of the $b$ branches.
The field which is left unmoved represents the original field that was on the link,
the others are the fields associated with the moved bonds within the interpretation of MK-RG as bond moving. 
The reason for this additional rule is commonly used and well explained in \cite{HL-pspin} to which we refer.
\section{RG flow and fixed points}
As discussed in the introduction, the RG flow is expected to be governed by the competition between the interactions and the 
external fields. This is indeed what we find. In order to discuss our results, it is useful to analyse the functional RG flow tracking the 
evolution of two special parameters: the variances of couplings and fields, denoted respectively $v_J(x)$ and $v_H(x)$ with  $x$ the iteration step and $\ell=2^x$ the corresponding length over which the RG is performed  (the temperature $T$ is an external tuning parameter).  
We first focus on the flow observed above the lower critical dimension, $d_L(M)$, which only depends on $M$ but it is otherwise the same
for all the models presented previously. For $d\ge d_L(M)$, at high enough temperature, $v_J(x)$ tends to zero and $v_H(x)$ to a constant, as expected for the liquid phase.  When lowering the temperature, 
both $v_J(x)$ and $v_H(x)$ initially grow, but after a certain scale $\xi=2^{x^*}$, $v_J(x)$ starts decreasing
and goes towards $0$, whereas $v_H(x)$ goes to a constant. The value of $x^*$, and hence the one of $\xi$, grows when lowering the temperature and diverges
at $T_c$. For $T<T_c$ the couplings and the external fields infinitely grow under renormalization and remain comparable on all scales: 
the system is in a glass phase which originates from the competition between interaction and idiosyncratic disorder.  
\vspace{-.9cm}
\begin{figure}[h]
\begin{center}
 \includegraphics[width=0.37\textwidth]{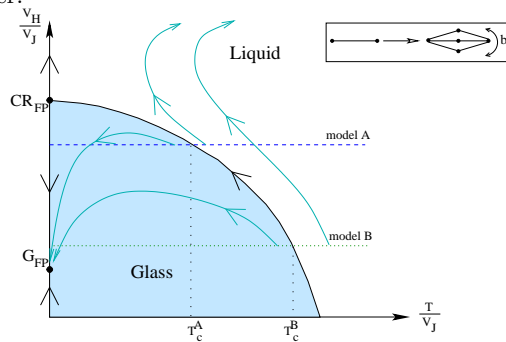}
\end{center}
\caption{Two dimensional (qualitative) projection of the flow diagram for $d\ge d_L$.
The two different dashed lines correspond to two different models. In the inset, basic step for the construction of a hierarchical lattice with $b=3$.}
\label{Fig:diagram}
\end{figure}
\vspace{-.5cm}
This RG flow is pictorially represented in Fig. \ref{Fig:diagram}: it is governed, in addition to the liquid fixed point ($L_{FP}$) placed at $(\frac{v_H}{v_J},\frac{T}{v_J})=(\infty,\infty)$, by the existence of two different fixed points corresponding to the critical regime ($CR_{FP}$) and to the low temperature glass phase ($G_{FP}$) respectively. 
Different models correspond to different initial conditions for the flow, as shown by the two dashed lines in Fig. \ref{Fig:diagram}. Accordingly, the value of the temperature $T_c$  at which the flow lies in the basin of attraction of $CR_{FP}$ is different. If the idiosyncratic disorder is too strong initially, i.e. $v_H/v_J$ is too large, no transition takes place. For all the three models we analysed, the initial ratio is low enough to induce a transition. Note that even if we put artificially $v_H=0$ at the first step, the external field is generated under RG\footnote{The case $M=1$ is an exception because of an additional (spin-flip) symmetry.  
In this case, there is an additional FP, placed at $(T/v_J,v_H/v_J)=(0,0)\equiv \text{SG}_{\text{FP}}$, 
that is stable on the line $v_H=0$ but unstable on the line $T=0$. For $M=1$, if we start with $v_H=0$, the flow goes to $\text{SG}_{\text{FP}}$. 
This result is natural since the case $M=1$ corresponds to a Spin Glass in field, that is
known to have a transition in zero field for $d\gtrsim2.5$ \cite{Dl_SG} and that was shown by us to display a MK-RG phase diagram with four fixed points in 
high enough dimension \cite{SGH_MK}.}.\\
The fixed points, $CR_{FP}$ and $G_{FP}$, are zero-temperature ones: both $v_J$ and $v_H$ increase with $\ell=2^x$ as $v_J(x)\propto\ell^{\theta}$, $v_H(x)\propto\ell^{\theta}$ in such a way that $v_H(x)/v_J(x)\rightarrow\text{constant}=(v_H/v_J)^*$. The values of $\theta$ and 
$(v_H/v_J)^*$ are universal for a given class of M-models in a given dimension: they only depend on $M$, $d$ and of course on the fixed point (we shall denote $\theta_{G}$ and $\theta_{CR}$ the values corresponding to $G_{FP}$ and $CR_{FP}$).\\
The glassy fixed point is attractive in all directions, whereas the critical one is attractive in all but one which heads toward either $G_{FP}$ or $L_{FP}$. The speed at which a small perturbation from the critical manifold grows determines the critical exponent $\nu$ of the correlation length $\xi\sim |T-T_c|^{-\nu}$. In order to compute $\nu$, we measure how two renormalized flows of the observable $v_H/v_J$ corresponding to different original $v_H$ and starting sufficiently near to $\text{CR}_\text{FP}$ distance themselves. The exponent $\nu$ only depends on $M$ and $d$ and grows for $d\rightarrow d_L(M)$. We show values of the exponents in Table 1. 
Note that since $\theta<d-1$ even for $G_{FP}$, the low temperature glass phase is non-trivial. As for spin-glasses, 
$\theta<d-1$ means that the system is critical, i.e. characterised by long-range critical correlations, even for $T<T_c$ \cite{SG_MK}. 
\vspace{-.8cm}
\begin{table}[h]
  \label{Tab:exponents}
\caption{Table of the critical properties for systems with different $M$ and $d$.
Values marked by * are very close to the lower critical dimension $d_L(M)$.}
  \begin{center}
  \begin{tabular}{| c | c || c | c | c | c | c |}
    \hline
 $M$  & $d$ & $\theta_{G}$ & $\theta_{CR}$ & $\nu$   \\
    \hline \hline
3 & 6.2 *  &  1.23714(7) & 1.0028(3) &   19.3(3)\\ \hline
3 & 6.6  & 2.17465(3) & 0.4624(1) &  3.32(6)\\ \hline
3 & 7.6  & 3.42715(1) & 0.1831(1) &   2.16(2)\\ \hline
4 & 5.7 *  & 1.27193(2) & 0.93985(2) &   12.1(1)\\ \hline
4 & 6.6  & 2.83068(1) & 0.266(2) &    2.54(4)\\ \hline
4 & 7.6  & 3.96445(2) & 0.117(1) &    1.76(2)\\ \hline
5 & 5.3 *  & 1.35347(5) & 0.84285(7) &  7.31(5)\\ \hline
  \end{tabular}
\end{center}

\end{table}
\vspace{-1cm}

\section{Lower critical dimension and avoided criticality}
We now focus on the dependence of $d_L(M)$ on $M$. We find that 
by increasing $M$ the lower critical dimension diminishes, i.e. glassiness is more robust 
for larger $M$s. This is also what found within mean-field theory and it is natural since 
for $M\rightarrow\infty$ one recovers the Kac-limit of mean-field models of glasses\footnote{Note that 
even for $M\rightarrow \infty$ the system is non mean-field like on large length-scales: the limits $M\rightarrow \infty$ and $\ell \rightarrow \infty$ 
for the RG flow do not commute. For large but finite $M$, the system is mean-field like on large but finite length-scales only; 
on even larger ones, the behaviour is non mean-field like and determined by the RG flow.}.
Note that large-$M$ systems could be more representative of glasses formed by interacting particles for which 
the degrees of freedom are continuous (corresponding naively to $M=\infty$).  \\

\begin{figure}[h]
\centerline{\includegraphics[width=.48\textwidth]{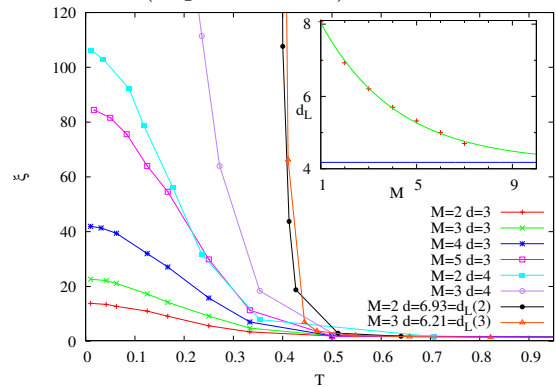}}
\vspace{-.5cm}
\caption{Correlation length (extracted as the length at which $v_J$ decays to half of its initial value) as a function of temperature for REM models with different $M$ and in different dimensions (temperatures are rescaled by $\sqrt{b}$, where $b$ is the number of branches of the HL lattice). If $d\ge d_L(M)$ the correlation length diverges below the critical temperature. If $d<d_L(M)$, instead, the 
correlation length initially grows but at a certain temperature it saturates to a constant value. Inset: 
Lower critical dimension for the $M$-value models as a function of $M$. The curve is the exponential function used to fit the data.
This fit gives an extrapolation for the lower critical dimension in the $M\rightarrow\infty$ limit to be $d_L(M=\infty)=4.18\pm0.14$. 
The straight line reports this extrapolation
}
\label{Fig:csi}
\end{figure}

In the inset of Fig. \ref{Fig:csi} the values of $d_L(M)$ are drawn as a function of $M$ up to $M=7$. 
By performing an exponential fit we obtain by extrapolation the value 
 $d_L(\infty)=4.18\pm0.14$. The prediction of MK-RG is therefore that three dimensional disordered spin models of glasses do not display a true glass transition. This however does not mean that they are not glassy. Indeed, we find a static length-scale $\xi$
that grows by lowering the temperature and eventually saturates to an $M$-dependent value. 
The saturation value of $\xi$ grows with $M$ and diverges when $d\uparrow d_L(M)$.  The main result of this analysis is that the growth of $\xi$ in three dimensions is due to the existence of $CR_{FP}$ and $G_{FP}$ fixed points in higher dimensions. This is a concrete realisation of the avoided transition scenario discussed previously (remember what happens to point D
of Fig \ref{Fig:Cartoon}): since the RG flow is regular in $d$, the existence of fixed points at $d_L$ influences the RG flow even at lower $d$ (the larger is $M$, the lower is $d_L$, the stronger is the influence). Our numerical analysis shows that for temperature close to the $T_c$ of the $d_L$-dimensional systems the probability distributions of $J,V_L$ and $V_R$ of the $d(<d_L)$-dimensional systems first approach 
the ones characteristic of $CR_{FP}$ for $d=d_L$ and then flow toward the ones characteristic of the liquid fixed point. 
Similarly, by lowering the temperature further, the RG flow is attracted more and more toward the vestige of the $G_{FP}$ and then eventually heads toward the liquid fixed point. This behaviour has to stop, and indeed does so, below a cross-over temperature since there is no real $G_{FP}$.  This is the mechanism that induces the growth of the
length-scale $\xi$ and hence, assuming that energy barriers scales as the interaction, is also the one that produces glassy and slow dynamics ($\xi$ corresponds to the length $\ell$ at which the flow starts heading toward $L_{FP}$). Note that the avoided transition mechanism plays an important role even for moderate values of $M$ for which $d_L$ is not close to three.
For example, for the finite dimensional REM models $M=3$, for which $d_L(3)\simeq 6.21$ we find a static length that grows from one to twentish lattice spacing in three dimensions, see Fig \ref{Fig:csi}. Similar results are obtained for the others $M-$value models.  
Since in experimentally relevant conditions dynamic length-scales grow at best from one 
to ten in molecular units and static ones are believed to be even smaller (but giving rise to extremely large times via the exponential relation $\tau\simeq \tau_0 e^{\Delta/T}$) \cite{OUP}, twenty is not at all a small value\footnote{Note that even if the static correlation length saturates to a finite value, if this is, say, larger than one hundred (in microscopic units) there would be absolutely no way to distinguish the avoided transition from a true one.}. In conclusion, the fixed points of the RG flow for $d=d_L$ could rule the three dimensional glassy behaviour observed in experiments and simulations even though $d_L$ is quite larger than three. 
Finally, let us point out that our MK-RG analysis suggests that the existence of a glass transition in three dimensions is a quantitative and not a qualitative issue: we do not find any fundamental reason to exclude the existence of an ideal glass transition in general. Indeed, it does take place if the dimension is high enough. The issue is just how low is the real value of $d_L(\infty)$ for which MK-RG only provides an approximated value (larger than three). 
\vspace{-.9cm}
\vspace{-.8cm}
\section{Relations with previous RG approaches}
With the aim 
of clarifying the state of the art and indicating new ways to proceed forward we now compare our findings and procedure to previous works on RG analysis of the glass transition. The one of 
\cite{CBTT} is the replica counterpart of ours, it consisted in performing the MK-RG directly 
on the replica field theory. It did find the existence of zero-temperature fixed points, but those were of simple nature: 
a discontinuity fixed point for the transition and a standard one for the low temperature phase, both with $\theta=d-1$. 
Moreover, the transition was found to persist down to $d=2$. We understand now that 
the saddle-point method used to solve the integral-RG equations is not fully justified. Physically, it misses 
the competition between the fluctuations of the interaction and of the idiosyncratic disorder crucial in our results \footnote{
Although the Boltzmann weight indeed increases and can be made very large from the very first RG iterations by taking $M$ large enough
the saddle-point method with a usual replica symmetry breaking structure used in \cite{CBTT} is unable to describe 
the disordered nature of the non-trivial zero temperature fixed points.}. It would be very interesting to understand how to take
this effect into account: this would allow one to recast our results in the replica field theory language and would open the way to the use of 
more refined and controlled non-perturbative field theory RG methods \cite{wetterich}. Another RG approach developed in recent years for glasses is the one based on Dyson hierarchical lattices. Although a full analysis of the corresponding RG equations has not been worked out yet, the results obtained for the finite dimensional REM do show the existence of a glass transition characterised by non-trivial critical exponents \cite{castellana}. It would be very interesting to go back to these studies and analyse whether the fixed points are zero-temperature ones, as predicted by our work. It was recently shown, both in the case of the RFIM \cite{rocchi} and of spin-glasses in a field  \cite{SGH_MK}, that Dyson RG has the potentiality of displaying such features.\\
The third important RG approach to glasses, which was developed by Moore and collaborators, is based on the same real space MK-RG method used in this work \cite{moore,mooredrossel}. They investigated the $M=2$ and $M=3$ $M-p$ models and found evidences
of avoided criticality. Furthermore, on the basis of their previous work \cite{mooredrossel} they claimed that the critical behaviour of disordered models of glasses is in the same universality class of spin-glasses in a field. Our findings confirm their results but the last one. The 
analysis presented in this work, which is based on a larger class of models and many different dimensions, reveals that the nature of the avoided critical point and the general structure of the RG flow is different from the one surmised in \cite{moore,mooredrossel}. In fact, there  is no spin-glass-like zero temperature fixed point as soon as $M>1$ in any dimension (compare and contrast the RG flow of Fig. \ref{Fig:diagram} to the one of spin-glasses in a field obtained in \cite{SGH_MK}).   
For small values of $M$, since the transition is strongly avoided and because of the proximity to the $M=1$ case, which is indeed spin-glass like, the explanation by Moore et al could indeed be valuable
\footnote{The numerical simulations of the $M=p=3$ model in three dimensions \cite{calta} indeed show a physics more spin-glass like.} but not for larger $M$, where glass-like physics sets in.    
\section{Discussion and conclusion}
We have presented a real space RG analysis of disordered models of glasses which are at the basis of  RFOT theory. It would be certainly interesting to complement our study by using another real space method, the Dyson Ensemble RG \cite{angeliniER}, which has the potentiality of being more accurate in high dimensions. In our study of spin-glasses in a field, corresponding to the $M=1$ case, we did it already and found
good agreement \cite{SGH_MK}. Even more interesting, but certainly more challenging, would be to develop a replica RG method 
able to connect the mean-field RFOT results to the RG ones. In fact, although the MK-RG provides a scenario that is compatible with RFOT theory \footnote{Note that the avoidance of a critical continuous transition leads to glassy phenomenology as explained in \cite{FLD}. Due to the approximate character of MK-RG we cannot conclude whether the transition found for $d>d_L$ is a purely continuous or mixed one, as in RFOT theory. We do expect that for large $d$, for which mean-field results become more reliable, it should be a mixed one.} key quantities  such as the configurational entropy or the overlap distribution 
between low temperature glass states cannot be probed directly but only inferred\footnote{This is a limitation of many real space methods. For instance, for MK-RG, once the renormalised length-scale has reached the system size, the number of possible low temperature states is at most $2^M\times 2^M$ hence making impossible to obtain the complex overlap distribution between states found within mean-field theory.

}. 
For instance, the behaviour of the specific heat at the transition (for $d>d_L$) is not discontinuous, which means that if the configurational entropy vanishes at $T_c$ \`a la Kauzmann, it does so  
more rapidly than linearly, as also found within the Dyson RG \cite{castellana}. 

In conclusion, our analysis unveils the existence and the nature of the RG fixed points responsible for glassy behaviour: two zero temperature fixed points, one for the transition and the other for the glass state.
It suggests that in three dimensions the glass transition is actually avoided but that glassiness is still driven by the RG fixed points present in higher dimensions. Given the approximate character of the MK-RG, there is some uncertainty on the true value of $d_L$. In order to 
firmly assess whether it is larger or smaller than three and to confirm and complement our results, it is crucial to develop 
a complete and more controlled RG analysis of the glass transition applicable both to spins and particles models. The present work 
provides the basis and the guidelines to face this challenge.  

\begin{acknowledgments}
We thank M. Moore, G. Tarjus and P. Urbani for discussions.
We acknowledge support from the ERC grants NPRGGLASS.  
\end{acknowledgments}



\end{article}

\end{document}